\documentclass[12pt,a4paper]{article}

\usepackage{jheplike,ifpdf,tocloft,rotating,amsmath,amssymb,mathtools,subcaption,xcolor}
\usepackage{hyperref,graphicx,array,multirow,color,mathrsfs,tikz}
\usetikzlibrary{calc,decorations.markings}

\widowpenalty=500\clubpenalty=1000\unitlength=1mm\textheight=8.7in\hypersetup{pdftitle={},pdfcreator={},linkcolor=[rgb]{0.15,0.35,0.75},colorlinks=true,citecolor=[rgb]{0.675,0,0.2},urlcolor=[rgb]{0.15,0.35,0.65}}
\let\olditemize\itemize\renewcommand{\itemize}{\vspace{-2pt}\olditemize\setlength{\itemsep}{1pt}\setlength{\parskip}{0pt}\setlength{\parsep}{-0pt}}
\let\oldenumerate\enumerate\renewcommand{\enumerate}{\vspace{-4pt}\oldenumerate\setlength{\itemsep}{1pt}\setlength{\parskip}{0pt}\setlength{\parsep}{0pt}}
\makeatletter\renewcommand\section{\addtocontents{toc}{\protect\addvspace{-2.25\p@}}\@startsection {section}{1}{\z@}{0.5ex \@plus .2ex \@minus 0.2ex}{0.3ex \@plus.1ex\@minus .5ex}{\normalfont\large\bfseries}}
\renewcommand\subsection{\addtocontents{toc}{\protect\addvspace{0.5\p@}}\@startsection {subsection}{1}{\z@}{0.5ex \@plus .2ex \@minus 0.2ex}{0.3ex \@plus.1ex\@minus .5ex}{\normalfont\bfseries}}
\renewcommand\subsubsection{\addtocontents{toc}{\protect\addvspace{-2.5\p@}}\@startsection {subsubsection}{1}{\z@}{0.5ex \@plus .2ex \@minus 0.2ex}{0.3ex \@plus.1ex\@minus .5ex}{\normalfont\bfseries}}
\newcommand{\fwbox}[2]{\text{\makebox[#1][c]{$\hspace{-150pt}\displaystyle#2\hspace{-150pt}$}}}
\newcommand{\fwboxL}[2]{\text{\makebox[#1][l]{$#2$}}}
\newcommand{\fwboxR}[2]{\text{\makebox[#1][r]{$#2$}}}
\newcommand{\bigger}[1]{\raisebox{-0.95pt}{\scalebox{1.25}{$#1$}}}
\newcommand{\Bigger}[1]{\raisebox{-2.25pt}{\scalebox{1.75}{$#1$}}}
\renewcommand{\bar}{\overline}
\renewcommand{\tilde}{\widetilde}
\newcommand{\eq}[1]{\vspace{-0.5pt}\begin{equation}\hspace{0pt}#1\hspace{-0pt}\vspace{-0.5pt}\end{equation}}
\newcommand{\fig}[2]{\vcenter{\includegraphics[scale=#1]{#2}}}
\newcommand{\mi}{\raisebox{0.75pt}{\scalebox{0.75}{$\hspace{-1pt}\,-\,\hspace{-0.75pt}$}}}
\newcommand{\pl}{\raisebox{0.75pt}{\scalebox{0.75}{$\hspace{-1pt}\,+\,\hspace{-0.75pt}$}}}
\newcommand{\ab}[1]{\langle #1\rangle}
\newcommand{\equivR}{\fwbox{14.5pt}{\hspace{-0pt}\fwboxR{0pt}{\raisebox{0.47pt}{\hspace{1.25pt}:\hspace{-4pt}}}=\fwboxL{0pt}{}}}
\newcommand{\equivL}{\fwbox{14.5pt}{\fwboxR{0pt}{}=\fwboxL{0pt}{\raisebox{0.47pt}{\hspace{-4pt}:\hspace{1.25pt}}}}}
\renewcommand{\u}[2]{(\hspace{-0.5pt}#1;\hspace{-1.5pt}#2\hspace{-0.5pt})}
\newcommand{\proj}[1]{\raisebox{1.75pt}{\big[}\hspace{-0.75pt}#1\hspace{-0.75pt}\raisebox{1.75pt}{\big]}}
\renewcommand{\r}[1]{\mathfrak{A}\hspace{-0.75pt}(\hspace{-1pt}{\color{hred}#1\hspace{-1pt}})}
\newcommand{\rb}[2]{(\hspace{-1pt}{\color{hblue}#1}\hspace{-1pt})^c_{\fwboxL{4.5pt}{#2}}}
\newcommand{\amp}[1]{\mathfrak{A}_{{\color{hred}#1}}}
\newcommand{\f}[1]{\mathfrak{#1}}
\newcommand{\Zeta}[1]{\zeta_#1}
\newcommand{\Li}[2]{\hspace{1pt}\mathrm{Li}_{#1}(#2)}

\newcommand{\polyalpha}{\rho}
\newcommand{\polybeta}{\sigma}

\definecolor{hblue}{rgb}{0,0,0.575}
\definecolor{hred}{rgb}{0.575,0.0,0.225}
\definecolor{hteal}{rgb}{0.0,0.545,0.7451}
\definecolor{optLegColour}{rgb}{0.5,0.5,0.5}

\def\figScale{0.835}\def\edgeLen{1*\figScale}\def\fScale{\small}\def\legLen{\edgeLen*0.65}\def\labelDist{\legLen*1.45}\def\lineThickness{(1pt)}\def\dotSize{(\figScale*10)}\def\ampSize{(1*\figScale*10pt)}\def\legSpread{5}\def\extLegLen{0.75*0.75*\figScale}
\tikzset{ddot/.style={fill=black,circle,minimum size=0.35*\dotSize,inner sep=0}}
\tikzset{int/.style={black,line width=\lineThickness,line cap=round,rounded corners=0.5pt}}\tikzset{ext/.style={black,line width=\lineThickness,line cap=round}}\tikzset{bdot/.style={fill=black,circle,minimum size=0.45*\ampSize,inner sep=0}}\tikzset{wdot/.style={draw=black,line width=\lineThickness,fill=white,circle,minimum size=0.65*\ampSize,inner sep=0}}
\newcommand{\leg}[3]{\draw[ext] #1--($#1+(#2:\legLen)$);\node at ($#1+(#2:\labelDist)$)[]{{\fScale #3}};}
\newcommand{\optLeg}[2]{\coordinate (aa0) at #1;\coordinate (aa5) at ($#1+(#2+\legSpread*3:\extLegLen)$);\coordinate (bb5) at ($#1+(#2-\legSpread*3:\extLegLen)$);\coordinate (aa1) at ($(aa0)!0.4!(aa5)$);\coordinate (aa2) at ($(aa0)!0.55!(aa5)$);\coordinate (aa3) at ($(aa0)!0.7!(aa5)$);\coordinate (aa4) at ($(aa0)!0.85!(aa5)$);\coordinate (bb1) at ($(aa0)!0.4!(bb5)$);\coordinate (bb2) at ($(aa0)!0.55!(bb5)$);\coordinate (bb3) at ($(aa0)!0.7!(bb5)$);\coordinate (bb4) at ($(aa0)!0.85!(bb5)$);\fill[optLegColour] (aa0)--(aa1)--(bb1)--(aa0);\fill[optLegColour] (aa2)--(aa3)--(bb3)--(bb2)--(aa2);\fill[optLegColour] (aa4)--(aa5)--(bb5)--(bb4)--(aa4);}

\newcommand{\octagonk}{\begin{tikzpicture}[scale=\figScale,baseline=-2.45]\useasboundingbox ($(-7.9/3*\figScale,-1.8)$) rectangle ($(7.9/3*\figScale,1.8)$);\draw[int,line width=0.1,red,draw=none] ($(-7.5/3*\figScale,-1.5)$) rectangle ($(7.5/3*\figScale,1.5)$);\coordinate(v1) at ($(0,0)+(90:\edgeLen/2)$);\coordinate(v2)at($(v1)+(18:\edgeLen)$);\coordinate(v3)at($(v2)+(18-72:\edgeLen)$);\coordinate(v4)at($(v3)+(18-2*72:\edgeLen)$);\coordinate(v5)at($(v4)+(18-3*72:\edgeLen)$);\coordinate(v6)at($(v5)+(198-0*72:\edgeLen)$);\coordinate(v7)at($(v6)+(198-1*72:\edgeLen)$);\coordinate(v8)at($(v7)+(198-2*72:\edgeLen)$);
\draw[int](v1)--(v2)--(v3)--(v4)--(v5)--(v6)--(v7)--(v8)--(v1);\draw[int](v1)--(v5);\leg{(v1)}{90}{8};\leg{(v2)}{72}{1};\leg{(v3)}{72-1*72}{\!2};\leg{(v4)}{72-2*72}{3};\leg{(v5)}{-90}{4};\leg{(v6)}{252-0*72}{5};\leg{(v7)}{252-1*72}{6\!};\leg{(v8)}{252-2*72}{7};\foreach\a in {1,3,5,7}{\node at (v\a) [bdot]{};};\foreach\a in {2,4,6,8}{\node at (v\a) [wdot]{};};
\node at ($(\edgeLen/1.4,0)$) []{$N_1$};\node at ($(-\edgeLen/1.4,0)$) []{$N_1$};
\end{tikzpicture}}
\newcommand{\octagonkPrime}{\begin{tikzpicture}[scale=\figScale,baseline=-2.45]\useasboundingbox ($(-7.9/3*\figScale,-1.8)$) rectangle ($(7.9/3*\figScale,1.8)$);\draw[int,line width=0.1,red,draw=none] ($(-7.5/3*\figScale,-1.5)$) rectangle ($(7.5/3*\figScale,1.5)$);\coordinate(v1) at ($(0,0)+(90:\edgeLen/2)$);\coordinate(v2)at($(v1)+(18:\edgeLen)$);\coordinate(v3)at($(v2)+(18-72:\edgeLen)$);\coordinate(v4)at($(v3)+(18-2*72:\edgeLen)$);\coordinate(v5)at($(v4)+(18-3*72:\edgeLen)$);\coordinate(v6)at($(v5)+(198-0*72:\edgeLen)$);\coordinate(v7)at($(v6)+(198-1*72:\edgeLen)$);\coordinate(v8)at($(v7)+(198-2*72:\edgeLen)$);
\draw[int](v1)--(v2)--(v3)--(v4)--(v5)--(v6)--(v7)--(v8)--(v1);\draw[int](v1)--(v5);\leg{(v1)}{90}{2};\leg{(v2)}{72}{3};\leg{(v3)}{72-1*72}{\!4};\leg{(v4)}{72-2*72}{5};\leg{(v5)}{-90}{6};\leg{(v6)}{252-0*72}{7};\leg{(v7)}{252-1*72}{8\!};\leg{(v8)}{252-2*72}{1};\foreach\a in {1,3,5,7}{\node at (v\a) [bdot]{};};\foreach\a in {2,4,6,8}{\node at (v\a) [wdot]{};};
\node at ($(\edgeLen/1.4,0)$) []{$N_1$};\node at ($(-\edgeLen/1.4,0)$) []{$N_1$};
\end{tikzpicture}}
\newcommand{\mhvInt}{\begin{tikzpicture}[scale=\figScale,baseline=-2.45]\useasboundingbox ($(-7.5/3*\figScale,-1.5)$) rectangle ($(7.5/3*\figScale,1.5)$);\draw[int,line width=0.1,red,draw=none] ($(-7.5/3*\figScale,-1.5)$) rectangle ($(7.5/3*\figScale,1.5)$);\coordinate(v1) at ($(0,0)+(90:\edgeLen/2)$);\coordinate(v2)at($(v1)+(18:\edgeLen)$);\coordinate(v3)at($(v2)+(18-72:\edgeLen)$);\coordinate(v4)at($(v3)+(18-2*72:\edgeLen)$);\coordinate(v5)at($(v4)+(18-3*72:\edgeLen)$);\coordinate(v6)at($(v5)+(198-0*72:\edgeLen)$);\coordinate(v7)at($(v6)+(198-1*72:\edgeLen)$);\coordinate(v8)at($(v7)+(198-2*72:\edgeLen)$);
\draw[int](v1)--(v2)--(v3)--(v4)--(v5)--(v6)--(v7)--(v8)--(v1);\draw[int](v1)--(v5);\optLeg{(v1)}{90};\leg{(v2)}{72}{$a$};\optLeg{(v3)}{72-1*72};\leg{(v4)}{72-2*72}{$b$};\optLeg{(v5)}{-90};\leg{(v6)}{252-0*72}{$c$};\optLeg{(v7)}{252-1*72};\leg{(v8)}{252-2*72}{$d$};\foreach\a in {1,3,5,7}{\node at (v\a) [bdot]{};};\foreach\a in {2,4,6,8}{\node at (v\a) [wdot]{};};
\node at ($(\edgeLen/1.4,0)$) []{$N_1$};\node at ($(-\edgeLen/1.4,0)$) []{$N_1$};
\end{tikzpicture}}

\title{\texorpdfstring{{\LARGE Rooting Out Letters: Octagonal Symbol}\\ {\LARGE Alphabets and Algebraic Number Theory}\\[-20pt]}{Rooting Out Letters: Octagonal Symbol Alphabets and Algebraic Number Theory}}
\author[a,b,c]{\vspace{-18pt}Jacob~L.~Bourjaily,}\emailAdd{bourjaily@psu.edu}
\author[a]{Andrew~J.~McLeod,}\emailAdd{amcleod@nbi.ku.dk}
\author[a]{Cristian~Vergu,}\emailAdd{c.vergu@nbi.ku.dk}
\author[a]{Matthias~Volk,}\emailAdd{mvolk@nbi.ku.dk}
\author[a]{Matt~von~Hippel,}\emailAdd{mvonhippel@nbi.ku.dk}
\author[a]{Matthias~Wilhelm}\emailAdd{matthias.wilhelm@nbi.ku.dk}
\affiliation[a]{Niels Bohr International Academy and Discovery Center, Niels Bohr Institute,\\University of Copenhagen, Blegdamsvej 17, DK-2100, Copenhagen \O, Denmark}
\affiliation[b]{Center for the Fundamental Laws of Nature, Department of Physics,\\ Jefferson Physical Laboratory, Harvard University, Cambridge, MA 02138, USA}
\affiliation[c]{Institute for Gravitation and the Cosmos, Department of Physics,\\Pennsylvania State University, University Park, PA 16892, USA}
\abstract{%
It is widely expected that NMHV amplitudes in planar, maximally supersymmetric Yang-Mills theory require symbol letters that are not rationally expressible in terms of momentum-twistor (or cluster) variables starting at two loops for eight particles. Recent advances in loop integration technology have made this an `experimentally testable' hypothesis: compute the amplitude at some kinematic point, and see if algebraic symbol letters arise. We demonstrate the feasibility of such a test by directly integrating the most difficult of the two-loop topologies required. This integral, together with its rotated image, suffices to determine the simplest NMHV component amplitude: the unique component finite at this order. Although each of these integrals involve algebraic symbol alphabets, the combination contributing to this amplitude is---surprisingly---rational. We describe the steps involved in this analysis, which requires several novel tricks of loop integration and also a considerable degree of algebraic number theory. We find dramatic and unusual simplifications, in which the two symbols initially expressed as almost ten million terms in over two thousand letters combine in a form that can be written in five thousand terms and twenty-five letters.
}
\preprint{}

\begin{document}
\maketitle

\vspace{\baselineskip}
\vspace{-0pt}\section{Introduction}\label{introduction_section}\vspace{-0pt}

The analytic structure and functional form of scattering amplitudes computed in (perturbative) quantum field theory continues to hold interesting surprises. Beyond leading order, amplitudes are typically transcendental functions---the simplest of which are known as generalized `polylogarithms': iterated integrals over differential forms with exclusively simple (logarithmic) poles in each integration variable. Although wider classes of functions are known to be needed for most amplitudes (see e.g.\ \cite{Broadhurst:1993mw,Laporta:2004rb,Bloch:2013tra,Adams:2013kgc,Bloch:2014qca,Adams:2016xah,Bloch:2016izu,Bourjaily:2019hmc,Bourjaily:2018yfy,Bourjaily:2018ycu,Bourjaily:2017bsb,Adams:2018bsn}), polylogarithms are often sufficient at low loop order and particle multiplicity, and are by far the best understood. Much of this understanding has emerged in the context of `symbology' \mbox{\cite{Goncharov:2010jf,Duhr:2011zq}}, which exploits the coproduct and Hopf algebra structure of these functions~\cite{Gonch2,Brown:2011ik,Brown1102.1312,Duhr:2012fh,Chavez:2012kn}. (See e.g.\ \cite{Duhr:2014woa} for an introduction to these ideas.)

One of the key aspects of symbols is that they encode complete information about the (iterated) branch cut structure of polylogarithms in terms of an \emph{alphabet} of primitive logarithmic branch-points called \emph{letters}. Knowledge about the alphabets relevant for certain polylogarithmic amplitudes has allowed incredible reaches into perturbation theory, well beyond what would be possible through any known (e.g.\ Feynman) diagrammatic expansion. Examples of such triumphs include the recent determination of certain six-particle amplitudes in planar maximally supersymmetric ($\mathcal{N}\!=\!4)$ Yang-Mills theory (sYM) through seven loops~\cite{Dixon:2011nj,Dixon:2013eka, Dixon:2014voa,Dixon:2014iba,Dixon:2015iva,Dixon:2016apl,Caron-Huot:2016owq,Caron-Huot:2019vjl,Caron-Huot:2019bsq}, and through four loops for seven particles \cite{Drummond:2014ffa,Dixon:2016nkn,Drummond:2018caf}. 

A microcosm of progress in scattering amplitudes more broadly, these calculations have fueled and been fueled by concrete examples. One still mysterious aspect of most known examples in this theory is that their symbol alphabets are found to be generated by \emph{cluster mutations}~\cite{Golden:2013xva}---rational transformations that define cluster algebras~\cite{Fomin:2003}. Such algebras naturally appear in the context of the positive Grassmannian geometry of on-shell scattering amplitudes~\cite{ArkaniHamed:2012nw}, and seem to encode physical aspects of amplitudes such as the Steinmann relations~\cite{Drummond:2017ssj,Drummond:2018dfd,Golden:2019kks,Mago:2019waa}; they also encode further types of structure whose physical interpretation remains less clear~\cite{Golden:2014pua,Golden:2014xqa,Golden:2018gtk}. 

Despite the intriguing role played by cluster algebras, it has long been known that even in planar sYM this story cannot be complete. Not only are non-polylogarithmic functions needed for most scattering amplitudes (at sufficiently high multiplicity or loop order), but even most polylogarithmic (N$^{k\geq2}$MHV) amplitudes at one loop require symbol letters that are not rationally related to any known cluster algebra. These algebraic roots arise, for example, as Gram determinants in the analysis of Landau singularities (see e.g.\ \cite{Dennen:2015bet,Dennen:2016mdk,Prlina:2017azl,Bourjaily:2018aeq}). 

It is therefore natural to wonder what kinds of letters arise in this theory's MHV and NMHV amplitudes, which have been argued to be polylogarithmic to all orders~\cite{Arkani-Hamed:2014via}. The symbol of all two-loop MHV amplitudes---computed in~\cite{CaronHuot:2011ky}---involve only letters drawn from the coordinates of Grassmannian cluster algebras (which are related to canonical coordinates on the space of positive momentum-twistor variables)~\cite{Golden:2013xva,Golden:2014pua}. Similarly, the symbol of the two-loop seven-point NMHV amplitude (computed in~\cite{CaronHuot:2011kk}) is entirely composed of cluster coordinates. Whether or not this continues to hold beyond seven particles constitutes an important open question. In particular, in~\cite{Prlina:2017azl} it was suggested that square roots could appear in NMHV amplitudes at two loops (and in MHV amplitudes at three loops) starting for eight particles. 

In this work, we probe the existence of these algebraic roots by directly computing a particular component of the eight-point two-loop NMHV amplitude. While we are not currently able to compute this component in full kinematics, it is sufficient to compute it analytically at a single (sufficiently generic) kinematic point. Note that it is, however, necessary to consider an entire amplitude, as it is well known that local integral representations can involve `spurious' symbol letters (or even `spurious' non-polylogarithmic parts---see e.g.\ \cite{Bourjaily:2015bpz,Bourjaily:2019iqr}) that cancel between terms. Surprisingly, in the component under study, this is precisely what happens: the local integrals that contribute to the amplitude individually involve quadratic roots, but these roots cancel. This of course has no implications for whether square roots will appear in other NMHV component amplitudes.

We begin in section~\ref{sec:main} by defining the particular component we will examine. In section \ref{subsec:directint}, we describe a direct integration strategy that can be used to compute it at a kinematic point; while it is not linearly reducible in the conventional sense, we find the integral can be divided up into parts that can be integrated after respective rationalizing changes of the integration variables. The resulting functional form involves many spurious algebraic letters in addition to the expected ones, so algebraic identities are required to eliminate them at symbol level, as we describe in section \ref{subsec:eliminating}. While the individual integrals contributing to this component contain quadratic roots, we show in section \ref{subsec:cancel} that the component as a whole does not. We then conclude, discussing further questions and potential applications. 

We also present two appendices. Appendix \ref{appendix:a_basis_for_nmhv_amplitudes} discusses a nice basis of R-invariants for this amplitude, while appendix \ref{sec:alg_num_th} reviews pertinent notions in algebraic number theory. We additionally include several ancillary files: the integrand of the integral we compute as \texttt{Omega1357Integrand.m}, expressions in multiple polylogarithms in \texttt{Omega1357MPLs.m} and \texttt{Omega3571MPLs.m}, and the simplified symbols we obtain as \texttt{Omega1357Symbol.m} and \texttt{Omega3571Symbol.m}. We also include a table of prime factorizations of the symbol letters conjectured in~\cite{Prlina:2017azl} for comparison with our results as \texttt{PrimeFactorLetters.pdf}.

\vspace{-0pt}\section{The Simplest NMHV Octagon Component Amplitude}\vspace{-0pt}\label{sec:main}

Explicit, prescriptive formulae for all two-loop $n$-point N$^k$MHV amplitude integrands for planar sYM, which we denote by $\mathcal{A}_n^{(k),2}$, were given in~\cite{Bourjaily:2015jna} (see also \cite{Bourjaily:2017wjl}); these amplitudes are expressed in terms of a basis of dual-conformal Feynman integrands involving only local propagators. Each integral in this basis can be Feynman parameterized and conformally regulated as described in~\cite{Bourjaily:2013mma,Bourjaily:2019jrk}. These integrals are not all yet known analytically. 

Consider for example the local integrand representation of MHV amplitudes at two loops \cite{ArkaniHamed:2010gh,ArkaniHamed:2010kv}:
\vspace{-2pt}\eq{\mathcal{A}_n^{(0),2}=\bigger{\displaystyle\sum_{\substack{1\leq a<b<c\\c<d<n+a}}}\!\!\mhvInt\,\equivL\bigger{\displaystyle\sum_{\substack{1\leq a<b<c\\c<d<n+a}}}\hspace{-4pt}\Omega[a,b,c,d]\,.\label{two_loop_mhv}\vspace{-0pt}}
Here, the `$N_1$'s indicate specific choices of loop-dependent numerators for these sets of (otherwise ordinary) Feynman propagators as defined in~\cite{Bourjaily:2015jna}.
Among these terms is the integral
\vspace{-0pt}\eq{\Omega\!\big[1,3,5,7\big]=\,\octagonk\;,\vspace{-0pt}\label{octagon_k_defined}}
which was referred to as `octagon K' in~\cite{Bourjaily:2018aeq}, where the particular challenges to its direct integration were described at some length (see also~\cite{Henn:2018cdp}). This integral is in fact the most difficult integral topology required for any eight-point amplitude at two loops for the simple reason that it is the only topology that depends on eight dual-momentum points. (Equivalently, it is the only topology which depends on 9 conformal degrees of freedom.) In general, the ratio function will involve all of the terms in (\ref{two_loop_mhv})---including $\Omega[1,3,5,7]$---because the 2-loop MHV amplitude is required to compute the ratio function. No analytic expression for $\Omega[1,3,5,7]$ is currently known, making the analysis of any octagon amplitude a considerable challenge.

Luckily, the question regarding whether or not algebraic letters appear in an amplitude can be answered for individual components. (We give a less component-oriented motivation for this amplitude in appendix~\ref{appendix:a_basis_for_nmhv_amplitudes}.) Moreover, provided the kinematics are parameterized appropriately, this question can be answered at a \emph{single kinematic point}. For the eight-point NMHV amplitude, there is in fact a \emph{simplest} component amplitude to consider:\footnote{Component fields of external supermultiplets are specified by their helicity and $SU(4)_R$-charges, written in superscript and subscript, respectively.}
\vspace{-1pt}\begin{align}
&\hspace{-75pt}\mathcal{A}_8\!\Big(\psi_1^{+\frac{1}{2}}\!\!,\phi^0_{12},\psi_2^{+\frac{1}{2}}\!\!,\phi^0_{23},\psi_3^{+\frac{1}{2}}\!\!,\phi^0_{34},\psi_4^{+\frac{1}{2}}\!\!,\phi^0_{41}\Big)&\label{finite_octagonal_component}\\
&\fwboxL{200pt}{\hspace{-14pt}=\int\!\!\!\big(
d\tilde{\eta}_8^{1}d\tilde{\eta}_1^{1}d\tilde{\eta}_2^{1}\big)\big(d\tilde{\eta}_2^{2}d\tilde{\eta}_3^{2}d\tilde{\eta}_4^{2}\big)\big(d\tilde{\eta}_4^{3}d\tilde{\eta}_5^{3}d\tilde{\eta}_6^{3}\big)\big(d\tilde{\eta}_6^{4}d\tilde{\eta}_7^{4}d\tilde{\eta}_8^{4}\big)\;\mathcal{A}_8\big(\lambda,\tilde\lambda,\tilde\eta\big)}\nonumber\\[-5pt]
&\fwboxL{200pt}{\hspace{-14pt}=\ab{82}\ab{24}\ab{46}\ab{68}\!\int\!\!\!\big(d\eta_1^1\big)\big(d\eta_3^2\big)\big(d\eta_5^3\big)\big(d\eta_7^4\big)\;\;\mathcal{A}_8^{}\big(\mathcal{Z}_1,\ldots,\mathcal{Z}_8\big)\,,}\nonumber
~\\[-22pt]\nonumber
\end{align}
where \mbox{$\ab{ab}\!\equivR\!\det\!\big(\lambda_a,\lambda_b\big)$} in terms of spinor variables with $p_a\!\equivL\lambda_a\smash{\tilde{\lambda}_a}$, and where $\eta_a$ is the fermionic component of the super momentum-twistor $\mathcal{Z}_a\equivR\!(z_a,\eta_a)$ \cite{Hodges:2009hk,Mason:2009qx,ArkaniHamed:2009vw}. This component amplitude is singled out by the fact that it happens to vanish exactly at tree level and one loop (see e.g.\ \cite{Bourjaily:2010wh,Bourjaily:2013mma,Bourjaily:2012gy}), rendering this two-loop amplitude \emph{infrared finite} and equal to the ratio function.  

Using the results of~\cite{Bourjaily:2015jna}, it is easy to confirm that the two-loop component (\ref{finite_octagonal_component}) in momentum-twistor variables is simply:
\vspace{1pt}
\eq{
\hspace{-290pt}\int\!\!\!d\eta_1^1d\eta_3^2d\eta_5^3d\eta_7^4\;\mathcal{A}_8^{L=2}=
\fwboxL{0pt}{\frac{1}{\ab{1357}}\!\!\left[\!\!\octagonk-\octagonkPrime\!\!\right],}\label{component_integrand}
}
where \mbox{$\ab{abcd}\!\equivR\!\det\!\big(z_a,z_b,z_c,z_d\big)$}. Notice that the \emph{sum} of these integrals contributes to the MHV amplitude (\ref{two_loop_mhv}), while their \emph{difference} is relevant to us here. The good news is that this component amplitude only requires one integral; the bad news is that it requires what is arguably the hardest eight-point integral at two loops.

Following~\cite{Bourjaily:2019jrk}, it is reasonably straightforward to Feynman parameterize (\ref{component_integrand}) without breaking conformal invariance. We give this Feynman-parametric representation in ancillary file \texttt{Omega1357Integrand.m}, expressed in terms of a particular momentum-twistor (cluster) coordinate chart (see~\cite{ArkaniHamed:2012nw,Bourjaily:2018aeq} for context):
\vspace{-0pt}\eq{\hspace{-150pt}Z\equivR\left(\begin{array}{@{}c@{$\;\;$}c@{$\;\;$}c@{$\;\;$}c@{$\;\;$}c@{$\;\;$}c@{$\;\;$}c@{$\;\;$}c@{}}\overline{s}_{23}&1&s_2s_3&0&\mi s_2s_3&0&\overline{s}_2s_3&0\\
\mi\overline{s}_3s_4&0&\overline{s}_{34}u&1&s_3 s_4&0&\mi s_3 s_4&0\\
s_1s_4&0&\mi s_1\overline{s}_4 u&0&\overline{s}_{41}u&1&s_1 s_4&0\\
\mi s_1s_2&0&s_1s_2 u&0&\mi \overline{s}_1s_2 u&0&\overline{s}_{12}u&1\end{array}\right)\;\Bigger{\Leftrightarrow}\fwboxL{0pt}{\hspace{-18pt}\fig{1}{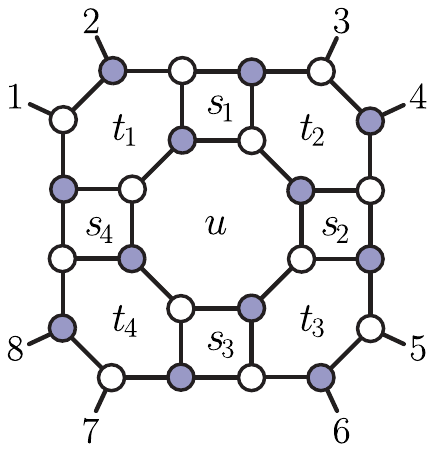}}\label{stu_cluster_chart}}
where $\overline{s}_{jk}\!\equivR\!(1\pl s_j\pl s_k\pl s_j s_k\pl  t_k)$ and $\overline{s}_i\!\equivR\!(1\pl s_i)$, introduced entirely for the sake of notational compression. Here, these coordinates correspond to the charts
\eq{
\fwbox{0pt}{\hspace{-25pt}s_1\equivR \frac{\ab{2346}\ab{4568}}{\ab{2468}\ab{3456}}\,,\;\; t_1\equivR\frac{\ab{1246}\ab{2345}\ab{3468}}{\ab{1234}\ab{2468}\ab{3456}}\,,\;\; u\equivR\frac{\ab{1248}\ab{2346}\ab{2678}\ab{4568}}{\ab{1246}\ab{2478}\ab{2568}\ab{3468}}\,,}}
with $s_2\equivR r^2(s_1)$, $t_2\equivR r^2(t_1)$, etc.\ defined by sequential two-fold rotations \mbox{$r^2\!\!:\!\!z_i\mapsto z_{i+2}$}. 

As described in~\cite{Bourjaily:2018aeq}, any rational parameterization of momentum twistors will be free of square roots associated with six-dimensional Gramians, and any rational point in momentum-twistor space can be accessed rationally in any cluster coordinate chart. And so the question of whether or not algebraic letters arise can be answered at \emph{any} rational point in momentum-twistor space. For the analysis described below, we chose to consider the (nearly symmetrical) point in kinematic space specified by the momentum-twistor matrix
\eq{Z\underset{}{\longrightarrow}Z^*\equivR \big(z_1,\ldots,z_n\big)\equivR\!\left(\begin{array}{@{}c@{$\;\;$}c@{$\;\;$}c@{$\;\;$}c@{$\;\;$}c@{$\;\;$}c@{$\;\;$}c@{$\;\;$}c@{}}5&1&1&0&\!\!\mi1\phantom{\mi}\!\!&0&2&0\\
\!\!\mi2\phantom{\mi}\!\!&0&5&1&1&0&\!\!\mi1\phantom{\mi}\!\!&0\\
1&0&\!\!\mi 2\phantom{\mi}\!\!&0&5&1&1&0\\
\!\!\mi 1\phantom{\mi}\!\!&0&1&0&\!\!\mi 2\phantom{\mi}\!\!&0&6&1\end{array}\right)\label{kinematic_point}}
obtained from (\ref{stu_cluster_chart}) by setting $t_2=2$ and all other coordinates $(s_i,t_i,u)$ to 1. Landau analysis (see \cite{Prlina:2017azl}) suggests that (\ref{octagon_k_defined}) may involve the roots associated with the four-dimensional Gramians:
\eq{\Delta\!\big[abcd\big]\equivR\sqrt{(1\mi u\mi v)^2\mi 4uv}\quad\text{with}\quad
u\equivR\u{ab}{cd}
\,,\quad
v\equivR\u{bc}{da}
\,,
}
where 
\eq{\u{ab}{cd}\equivR\displaystyle\!\frac{\ab{a-1a\,b-1b}\ab{c-1c\,d-1d}}{\ab{a-1a\,c-1c}\ab{b-1b\,d-1d}}\,.} 
For the kinematic point defined by (\ref{kinematic_point}), these are
\eq{\Delta\!\big[1357\big]=\frac{1}{806}\sqrt{644801}\,,\quad\Delta\!\big[2468\big]=\frac{1}{5}\sqrt{21}\,.\label{expected_square_roots}}
Our question, therefore, is whether or not the roots (\ref{expected_square_roots})---or any others---arise as part of the symbol alphabet for the component (\ref{component_integrand}). Answering this question turned out to require more cleverness and subtlety than expected. 
We shall now describe the concrete steps involved.

\vspace{-0pt}\subsection{Direct, (Feynman-)Parametric Integration of \texorpdfstring{$\Omega[1,3,5,7]$}{Omega[1,3,5,7]}}\vspace{-0pt}
\label{subsec:directint}

The loop-momentum integral over $\Omega[1,3,5,7]$ corresponds to a five-fold parametric integral of Feynman (or Schwinger) parameters:
\eq{\hspace{-5pt}\Omega\big[1,3,5,7\big]\equivL\int\limits_0^\infty\!\!\!\proj{d^3\vec{\alpha}}d^2\vec{\beta}\;\;\mathcal{I}(\alpha_1,\ldots,\alpha_4,\beta_1,\beta_2)\,.\label{Feynman_param_int}}
Here, the integrals over $\{\alpha_1,\ldots,\alpha_4\}$ are projective, and those over $\beta_1,\beta_2$ are not. (This distinction is irrelevant from the viewpoint of the Cheng-Wu theorem, but reflects how the parameterization was derived.) 

The principle obstruction to parametric integration is that $\mathcal{I}(\vec{\alpha},\vec{\beta})$ is not linearly reducible in the sense of~\cite{Brown_2009}.
In particular, using compatibility-graph reduction \cite{Brown:2009ta} (as implemented for example in the package \texttt{HyperInt} \cite{Panzer:2014caa}), one can readily find that at most two integrations can be carried out without introducing algebraic roots. 
For instance, upon integrating out $\beta_1$ and $\beta_2$ (in that order), further 
integration seems to be obstructed along every path. For example, the pathway in which $\alpha_1$ is integrated next is obstructed by the existence of a quadratic polynomial $Q_1(\alpha_1)$ in the denominator, as this leads to a result that involves the square root of the discriminant of $Q_1$; this square root involves the remaining integration parameters, na\"ively taking us out of the space of multiple polylogarithms. There is a similar obstruction with respect to $\alpha_4$, due to a quadratic denominator factor $Q_4(\alpha_4)$. (The obstructions in $\alpha_2$ and $\alpha_3$ are given by three quadratic polynomials each.)

Luckily, after integrating over $\beta_1$ and $\beta_2$ , there are no terms that \emph{simultaneously} depend on both quadratic factors $Q_1(\alpha_1)$ and $Q_4(\alpha_4)$. Thus, we may divide them according to whether or not $Q_1(\alpha_1)$ appears. Specifically, we define
\eq{\int\limits_0^\infty\!\!\!d^2\vec{\beta}\;\;\mathcal{I}(\alpha_1,\ldots,\alpha_4,\beta_1,\beta_2)\equivL\mathcal{I}(\vec{\alpha})\equivL\; \mathcal{I}_A+\mathcal{I}_B\,,}
with $\mathcal{I}_B$ consisting of all terms that involve $Q_1(\alpha_1)$, and $\mathcal{I}_A$ being all terms that do not depend on $Q_1(\alpha_1)$. To be clear, $\mathcal{I}_A$ consists of both those terms involving $Q_4(\alpha_4)$, and also those depending on neither quadratic factor.
Note that $\mathcal{I}_A$ and $\mathcal{I}_B$ are separately finite.

Before we describe further integrations, it is worth mentioning one potential subtlety. We will ultimately be interested in fixing the projective redundancy of different parts of the original integral in different ways. To do so, we must first reprojectivize these integrals by making the replacement $\alpha_i\mapsto\alpha_i/(\sum\alpha_i)$.\footnote{This is due to the arguments of the logarithms (and polylogarithms) introduced by previous integrations, which are not homogeneous.} This is done before we set any parameter to unity. 

\begin{figure}[t]\eq{\begin{tikzpicture}[scale=\figScale,baseline=-2.45]\useasboundingbox ($(-7*\figScale,-2.85)$) rectangle ($(7*\figScale,2.85)$);\draw[int,line width=0.1,red,draw=none] ($(-11*\figScale,-2.85)$) rectangle ($(11*\figScale,2.85)$);\node (v0) at (-9.25,0) [anchor=west] {$\mathcal{I}\big(\vec{\alpha},\vec{\beta}\big)\!\!\!$};\node (v1) at (-5,0) [] {$\mathcal{I}\big(\vec{\alpha}\big)\!\equivL\!\!\left\{\rule{0pt}{50pt}\right.$};\node (v1a) at(-4.125,1.5) [anchor=west] {$\!\mathcal{I}_{\fwbox{5pt}{A}}\big[\!\!\not\owns\!\! Q_1(\alpha_1)\!\big]\!\!$};\node (v1b) at(-4.125,-1.5) [anchor=west] {$\!\mathcal{I}_{\fwbox{5pt}{B}}\big[\!\!\owns\!\! Q_1(\alpha_1)\!\big]\!\!$};\node (v2a0) at(1.725,2.25) [anchor=west] {$\!\!\!\mathcal{I}_{\fwbox{8pt}{A_0}}\big[\!\!\not\owns\!\! \sqrt{q_1},\sqrt{q_2}\!\big]$};\node (v3a0) at(8.125,2.25) [anchor=west] {$\!\!I_{\fwbox{8pt}{A_0}}$};\node (v3a1) at(8.125,1.5) [anchor=west] {$\fwboxR{0pt}{\left\{\rule{0pt}{30pt}\right.}\!\!I_{\fwbox{8pt}{A_1}}$};\node (v3a2) at(8.125,0.75) [anchor=west] {$\!\!I_{\fwbox{8pt}{A_2}}$};\node (v2a1) at(1.725,1.5) [anchor=west] {$\!\fwboxR{0pt}{\left\{\rule{0pt}{30pt}\right.}\!\!\mathcal{I}_{\fwbox{8pt}{A_1}}\big[\!\!\owns\!\! \sqrt{q_1(\alpha_3,\alpha_4)}\!\big]\!\fwboxL{0pt}{\left.\rule{0pt}{30pt}\right\}}\hspace{7.5pt}$};\node (v2a2) at(1.725,0.75) [anchor=west] {$\!\!\mathcal{I}_{\fwbox{8pt}{A_2}}\big[\!\!\owns\!\! \sqrt{q_2(\alpha_3,\alpha_4)}\!\big]$};\node (v2b0) at(1.725,-0.75) [anchor=west] {$\!\!\mathcal{I}_{\fwbox{8pt}{B_0}}\big[\!\!\not\owns\!\! \sqrt{\tilde{q}_1},\sqrt{\tilde{q}_2}\!\big]$};\node (v2b1) at(1.725,-1.5) [anchor=west] {$\!\fwboxR{0pt}{\left\{\rule{0pt}{30pt}\right.}\!\!\mathcal{I}_{\fwbox{8pt}{B_1}}\big[\!\!\owns\!\! \sqrt{\tilde{q}_1(\alpha_1,\alpha_2)}\!\big]\!\fwboxL{0pt}{\left.\rule{0pt}{30pt}\right\}}\hspace{7.5pt}$};\node (v2b2) at(1.725,-2.25) [anchor=west] {$\!\!\mathcal{I}_{\fwbox{8pt}{B_2}}\big[\!\!\owns\!\! \sqrt{\tilde{q}_2(\alpha_1,\alpha_2)}\!\big]$};\node (v3b0) at(8.125,-0.75) [anchor=west] {$\!\!I_{\fwbox{8pt}{B_0}}$};\node (v3b1) at(8.125,-1.5) [anchor=west] {$\fwboxR{0pt}{\left\{\rule{0pt}{30pt}\right.}\!\!I_{\fwbox{8pt}{B_1}}$};\node (v3b2) at(8.125,-2.25) [anchor=west] {$\!\!I_{\fwbox{8pt}{B_2}}$};\draw [int,->] (v0) to (v1);\draw [int,->] (v1a) to ($(v2a1)-(2.5,0)$);\draw [int,->] (v1b) to ($(v2b1)-(2.5,0)$);\draw [int,->] (v2a1) to ($(v3a1)-(0.8,0)$);\draw [int,->] (v2b1) to ($(v3b1)-(0.8,0)$);\node at ($($(v0)-(0.0,0)$)!.5!(v1)$) [anchor=south] {$\hspace{-4pt}\displaystyle\int\!\!\!d^2\vec{\beta}$};\node at ($(v1a)!.5!($(v2a1)-(0.4,0)$)$) [anchor=south] {$\hspace{-13pt}\displaystyle\int\!\!\!d\alpha_1,\!\!\int\!\!\!d\alpha_2$};\node at ($(v1b)!.5!($(v2b1)-(0.4,0)$)$) [anchor=south] {$\hspace{-13pt}\displaystyle\int\!\!\!d\alpha_4,\!\!\int\!\!\!d\alpha_3$};\node at ($(v2a1)!.5!($(v3a1)-(0.0,0)$)$) [anchor=south] {$\hspace{30pt}$``$\!\!\displaystyle\int\!\!\!d\alpha_3$''};\node at ($(v2a1)!.5!($(v3a1)-(0.0,0)$)$) [anchor=north] {$\hspace{28pt}\alpha_4\!\to\!1$};\node at ($(v2b1)!.5!($(v3b1)-(0.0,0)$)$) [anchor=south] {$\hspace{30pt}$``$\!\!\displaystyle\int\!\!\!d\alpha_1$''};\node at ($(v2b1)!.5!($(v3b1)-(0.0,0)$)$) [anchor=north] {$\hspace{28pt}\alpha_2\!\to\!1$};\end{tikzpicture}\nonumber\vspace{-10pt}}\caption{Integration strategy for  $\Omega[1,3,5,7]$. Here, the final integrations are written in quotes to clarify that this step should be understood as integration \emph{after} the changes of variables made to rationalize the quadratic roots; these changes depend on which roots exist, and so are different for different groups $\mathcal{I}_{A_i}$ and $\mathcal{I}_{B_i}$.}\label{integration_strategy_figure}\end{figure}

Let us first consider the integration of $\mathcal{I}_A$. Free of the quadratic obstruction $Q_1(\alpha_1)$, we can integrate over $\alpha_1$ and subsequently $\alpha_2$, leaving us with a one-fold projective integral. The $\alpha_2$ integration, however, result in terms that involve square roots of two more irreducible quadratics $q_1(\alpha_3,\alpha_4)$ and $q_2(\alpha_3,\alpha_4)$. While the appearance of such factors would generally obstruct further integration, it turns out that no single term contains both roots. Thus, we can further divide $\mathcal{I}_A$ into three parts: $\mathcal{I}_{A_0}$, which is free of any square roots, 
$\mathcal{I}_{A_1}$, which involves only $\sqrt{q_1(\alpha_3,\alpha_4)}$, and $\mathcal{I}_{A_2}$, which involves only $\sqrt{q_2(\alpha_3,\alpha_4)}$. After setting the projective variable $a_4=1$,  we can then use a standard change of variables known as Euler substitution (see also \cite{Besier:2018jen}) to rationalize $\sqrt{q_1(\alpha_3,1)}$ and $\sqrt{q_2(\alpha_3,1)}$, respectively, in the latter two groups.

We can integrate each of the terms in $\mathcal{I}_B$ following a very similar strategy. Specifically, we first integrate out $\alpha_4$ and then $\alpha_3$, which results in terms that individually involve one (or neither) of a pair of square roots of different quadratic polynomials, $\tilde{q}_1(\alpha_1,\alpha_2)$ and $\tilde{q}_2(\alpha_1,\alpha_2)$. Splitting these pieces in the same way as for $\mathcal{I}_A$, fixing $\alpha_2=1$ and changing variables to rationalize each root, we can do the final integration.

The steps involved in this divide-and-conquer strategy are summarized in \mbox{figure \ref{integration_strategy_figure}}. The result is a sum of terms, each expressed in terms of multiple polylogarithms depending on algebraic arguments of high degree (up to 16 in some cases). These expressions can be evaluated to arbitrarily high precision---for example, using \texttt{GiNaC} \cite{Bauer:2000cp,Vollinga:2004sn}---and have been checked to agree with the numerical (Monte Carlo) integration of the Feynman parametric integral (in {\sc Mathematica}). We attach these results as \texttt{Omega1357MPLs.m} and \texttt{Omega3571MPLs.m}.

Unfortunately, as mentioned, the multiple polylogarithms that arise in this process depend on \emph{many} algebraic roots. In addition to the expected roots from the Landau analysis at this kinematic point, $\sqrt{21}$ and $\sqrt{644801}$, we find that $\Omega\big[1,3,5,7\big]$ and $\Omega\big[3,5,7,1\big]$ each involve 85 distinct square roots, with only 12 in common between the two integrals. Each also involves irreducible roots of four distinct fourth-order polynomials, only one of which appears in both integrals. The vast majority of these algebraic roots are certain to be `spurious': arising entirely through the change of variables introduced in the final stages of the integration strategy (required to rationalize the final integrations). 
To assess whether or not these roots (or any others) are truly spurious, we analyze the symbol of each integral.

\vspace{-0pt}\subsection{Eliminating Identities Among `Spurious' Algebraic Letters}\vspace{0pt}\label{subsec:eliminating}

As described above, we are able to evaluate $\Omega\big[1,3,5,7\big]$ and $\Omega\big[3,5,7,1\big]$ as complicated expressions in terms of multiple polylogarithms, which we expect to satisfy many nontrivial relations. To investigate these relations, we take the symbol of each function.\footnote{
It is sometimes colloquially stated that the symbol of a constant is zero; while this is true for the constants we most familiarly encounter (namely, the multiple zeta values), it is not true in general. The only two letters that are dropped in the symbol are $\pm 1$ (which correspond to $\log(1)=0$ and $i \pi$).}
Doing so, we obtain a pair of extremely complicated expressions, each involving a large number of spurious letters. Factoring each letter na\"ively (including factoring any integer primes), $\Omega\big[1,3,5,7\big]$ has a symbol composed of 8,367,616 terms that involve 2,024 letters, while the symbol of $\Omega\big[3,5,7,1\big]$ contains 9,941,483 terms and 2,156 letters.

Clearly, these symbols must be simplified. To do so, we want to find a set of
multiplicatively independent letters \(\mathcal{S}\) in terms of which both of these symbols can be expressed. 
Landau analysis suggests that the final alphabet $\cal S$ should be drawn from the union of the two algebraic number fields
\(\mathbb{Q}(\sqrt{21})\bigger{\cup}\mathbb{Q}(\sqrt{644801})\).
However, our integration procedure yields a symbol with a \emph{much} larger initial alphabet, involving for instance algebraic numbers up to degree \(16\). Finding algebraic relations between these complicated letters in order to reduce them to elements of \(\mathcal{S}\) can be extremely difficult. To give the reader a sense of this complexity, we consider some examples. 

Let \(P_i\!\in\! K[X]\) be some degree-four polynomials (indexed by $i$) with coefficients\footnote{To be more precise, two of these minimal polynomials are with coefficients in \(\mathbb{Z}\), one is with coefficients in \(\mathbb{Z}[\sqrt{21}]\) while another has coefficients in \(\mathbb{Z}[\sqrt{644801}]\).} in \(K = \mathbb{Q}(\sqrt{21}, \sqrt{644801})\). Our initial alphabet includes various roots of $P_i$, denoted $\polybeta_{i,r}^*$ for $r=1,\ldots,4$. An example of the kind of roots that arise for us are those of the fourth-degree polynomial:
\begin{align}
P_1 =&\phantom{\,+}(515426609 + 641880 \sqrt{644801}) +  (2105546840 + 2622160 \sqrt{644801}) X\nonumber\\
&+\!(3225674840 + 4015200 \sqrt{644801}) X^2 +  (2240256000 + 2676800 \sqrt{644801}) X^3\nonumber\\
&+\!1120128000 X^4\,.\label{bad_roots_arising_from_bad_polys}
\end{align}
Clearly, we expect the four roots of $P_1$ that arise in our symbol alphabet  to be spurious. Therefore, we must find some way to demonstrate that they cancel. 

Actually, an alphabet merely involving $\polybeta_{i,r}^*$ would not be so difficult. It turns out in our case that the most complicated letters we see are of the type
\(\polyalpha - \polybeta^*_{i, r}\), where \(\polyalpha\) can be an integer or a
linear combination of up to two square roots.  When there are two
roots, one always belongs to \(K\).  Furthermore, when
\(\polyalpha = m + n \sqrt{c}\) with \(m, n \in K\) and \(c \in \mathbb{Z}\)
appears, then its conjugate
\(\bar{\polyalpha} = m - n \sqrt{c}\) also appears.

There are two types of relations involving the roots \(\polybeta_{i,r}^*\) that
turn out to be useful for us.  The first type involves products
\(\prod_{r = 1}^4 (\polyalpha - \polybeta^*_{i, r})\).  These
products are completely symmetric in the roots of \(P_i\), so they
belong to an extension of the field \(K\) by \(\polyalpha\)---in
particular, they can be written as linear combinations of square roots and integers. 
Actually, it turns out that products of certain pairs of roots of
\(P_i\) also yield simple answers.  We believe it should be possible to
explain the existence of these latter mysterious identities using Galois
theory, but we have not performed this analysis.

The second type of identities involve products of type
\((\polyalpha - \polybeta^*_{i, r})(\bar{\polyalpha} - \polybeta^*_{i, r})\), where
\(\bar{\polyalpha}\) is one of the conjugates of \(\polyalpha\).  Expanding
out this product we obtain a degree-two polynomial in
\(\polybeta^*_{i, r}\) with coefficients in \(K\).  Next, we search for
exponents \(e_\polyalpha\) corresponding to values of \(\polyalpha\) such
that, in the product of these letters raised to power \(e_\polyalpha\), the
\(\polybeta^*_{i, r}\) cancels and the answer is of degree two.  It turns
out to be sufficient to bound the search so that
\(\lvert e_\polyalpha\rvert \leq 2\).  The calculation of these products
can be conveniently performed using \texttt{SageMath}~\cite{sagemath}, which uses
\texttt{Pari}~\cite{PARI2}.

Let us be more concrete with an example of this second type of identity. For the polynomial $P_1$ given in (\ref{bad_roots_arising_from_bad_polys}), we find the letters
\eq{\begin{split}
\hspace{-20pt}a_1(\polybeta_{1,r}^*)\equivR&(1668888 +2080 \sqrt{644801}) +(25600\polybeta_{1,r}^* + 4160 \sqrt{644801}) \polybeta_{1,r}^*,\hspace{-50pt}\\
\hspace{-20pt}a_2(\polybeta_{1,r}^*)\equivR&(1412136 + 1760 \sqrt{644801}) + (3097600\polybeta_{1,r}^* + 3520 \sqrt{644801}) \polybeta_{1,r}^*,\hspace{-50pt}\\
\hspace{-20pt}a_3(\polybeta_{1,r}^*)\equivR&(10013328 + 12480 \sqrt{644801}) + (17305600\polybeta_{1,r}^* + 24960 \sqrt{644801}) \polybeta_{1,r}^*,\hspace{-50pt}\\
\hspace{-20pt}a_4(\polybeta_{1,r}^*)\equivR&(11938968 + 14880 \sqrt{644801}) + (24601600\polybeta_{1,r}^* + 29760 \sqrt{644801}) \polybeta_{1,r}^*,\hspace{-50pt}\\
\hspace{-20pt}a_5(\polybeta_{1,r}^*)\equivR&(2456474760 + 3061600 \sqrt{644801}) + (5069440000\polybeta_{1,r}^* + 6123200 \sqrt{644801}) \polybeta_{1,r}^*\hspace{-50pt}
\end{split}}
(among many others involving $\polybeta_{1,r}^*$), where $\polybeta_{1,r}^*$ is any root of $P_1$. It is not hard to verify that
\eq{\frac{a_1a_2^2}{a_3a_4a_5}=-\frac{121}{358670}\in K\,}
using \texttt{SageMath} (or even {\sc Mathematica}).

Fortunately, the method described above turns out to be sufficient to
find all required relations between the most complicated letters that appear in our initial symbols, allowing us to get rid of all higher-degree roots.  
However, many other potentially-spurious letters
remain---in particular, there still exist linear combinations of
up to two square roots, and square roots beyond the two physical ones in~\eqref{expected_square_roots}.  

For the letters containing square roots, we group them according to the
algebraic number fields to which they belong and compute the
factorization of the principal ideal they generate (see appendix~\ref{sec:alg_num_th} for more details).  For this step we
use again \texttt{SageMath} and \texttt{Pari}.  Using this
factorization, we can find multiplicative relations between these letters.  Note
that the integer prime factors we generated in the first step belong to
each of these number fields, so their decomposition in prime ideals has to
be computed as well.

This factorization also contains a unit part, which is a term belonging
to the group of units of the various rings we consider.  In some of the cases we encounter, the unit part is \(\pm 1\), but
in others it is non-trivial.  We keep a list of all the units arising
during the calculations in a given ring, and if two of them are
identical we obtain a new identity by taking the ratio.  In principle
a more sophisticated approach is possible.

Using these methods, we decompose our letters into a multiplicatively independent set $\mathcal{S}$.  Doing so, many of the spurious letters in our symbols combine cleanly into integer letters. Others cancel entirely, removing terms and causing other spurious letters to drop out. In the end, we find the symbol of each function simplifies dramatically. Expressing $\Omega\big[1,3,5,7\big]$ and $\Omega\big[3,5,7,1\big]$ in terms of a shared, multiplicatively independent symbol alphabet, we find only 35 letters are needed. These letters only involve the expected square roots: five involve $\sqrt{644801}$, two involve $\sqrt{21}$, and the rest are integer primes. Expressed in these letters, $\Omega\big[1,3,5,7\big]$ is 5316 terms long, while $\Omega\big[3,5,7,1\big]$ contains 5245 terms. We attach the symbol of each in ancillary files \texttt{Omega1357Symbol.m} and \texttt{Omega3571Symbol.m}, respectively.

Interestingly, some of the symbol letters that contain $\sqrt{21}$ and $\sqrt{644801}$ can be constructed simply in dual twistor space. Namely, out of eight points $z_1,\dots,z_8$, we can form four skew lines $(z_1,z_2)$, $(z_3,z_4)$, $(z_5,z_6)$, $(z_7,z_8)$. These four skew lines have two transversals (lines that intersect all four of them). From the points of intersection on each of these transversals we can form a cross ratio. A similar construction can be carried out starting from the $(z_2,z_3)$, $\dots$, $(z_8,z_1)$. Some of the cross ratios that can be formed in this way appear directly appear in our symbol expression for $\Omega\big[1,3,5,7\big]$ and $\Omega\big[3,5,7,1\big]$.

\subsection{Cancellations in the Component Amplitude}\label{subsec:cancel}

Individually, $\Omega\big[1,3,5,7\big]$ and $\Omega\big[3,5,7,1\big]$ both contain square-root letters. Now that we have expressed them in the same alphabet, we can examine their difference $\Omega\big[1,3,5,7\big]-\Omega\big[3,5,7,1\big]$, the combination that appears in this component of the NMHV amplitude. Remarkably, this difference is free of square-root letters! Recall that we are using a multiplicatively independent alphabet: as such, the vanishing of square roots in $\Omega\big[1,3,5,7\big]-\Omega\big[3,5,7,1\big]$ requires that terms involving each of the six independent square-root-containing letters cancel separately. We find that the difference $\Omega\big[1,3,5,7\big]-\Omega\big[3,5,7,1\big]$ contains just 25 letters, all integer primes.

The sum $\Omega\big[1,3,5,7\big]+\Omega\big[3,5,7,1\big]$ contributes to the eight-point MHV amplitude. This sum is not free of square roots, and depends on all of the letters present in the two integrals. This observation is still consistent with the observed absence of square roots in the alphabet of the two-loop eight-point MHV amplitude because several other root-containing integrals contribute to this amplitude---including two other permutations of the integral we computed here. Other cancellations, much like those we observed, must be present in this combination.

We find that square-root letters are present in the second and third entry of $\Omega\big[1,3,5,7\big]$ and $\Omega\big[3,5,7,1\big]$, but not the first or fourth entry. This is as expected, as first entries should correspond to Mandelstam invariants while last entries are constrained by the $\bar{Q}$ equation~\cite{CaronHuot:2011kk}. More specifically, first entries should be composed of four-brackets of the form $\ab{i,i+1,j,j+1}$. Examining our symbol, we find first entries of $2, 3, 5, 11, 13$, and $31$. Computing the expected first entries at our kinematic point, we find
\eq{\begin{split}
\ab{1, 2, 3, 4}&=1\,,\,\,\,\,\quad\ab{1, 2, 4, 5}=3\,,\,\,\,\,\quad
\ab{1, 2, 5, 6}=5\,,\,\,\,\,\quad\ab{1, 2, 6, 7}=13\,,\\
\ab{1, 2, 7, 8}&=1\,,\,\,\,\,\quad\ab{2, 3, 4, 5}=1\,,\,\,\,\,\quad
\ab{2, 3, 5, 6}=11\,,\quad\ab{2, 3, 6, 7}=31\,,\\
\ab{2, 3, 7, 8}&=3\,,\,\,\,\,\quad\ab{1, 2, 3, 8}=1\,,\,\,\,\,\quad
\ab{3, 4, 5, 6}=1\,,\,\,\,\,\,\quad\ab{3, 4, 6, 7}=4\,,\\
\ab{3, 4, 7, 8}&=5\,,\,\,\,\,\quad\ab{1, 3, 4, 8}=11\,,\quad
\ab{4, 5, 6, 7}=2\,,\,\,\,\,\quad\ab{4, 5, 7, 8}=11\,,\\
\ab{1, 4, 5, 8}&=26\,,\quad\ab{5, 6, 7, 8}=1\,,\,\,\,\,\quad
\ab{1, 5, 6, 8}=3\,,\,\,\,\,\quad\ab{1, 6, 7, 8}=1\,,
\end{split}
}
which indeed cover all observed first entries. 

We can also investigate whether the prime-number symbol entries we observe elsewhere in the symbol can originate from the entries predicted in~\cite{Prlina:2017azl}. We have attached this analysis as ancillary file \texttt{PrimeFactorLetters.pdf}, where we tabulate the prime factors of each of the predicted letters at this kinematic point. We find these factors span all of the letters that we observe. There are additional prime factors occurring in predicted letters in~\cite{Prlina:2017azl} that we do not observe; these are marked by an asterisk in our table.

In addition to these observations, we find that the two square roots $\sqrt{644801}$ and $\sqrt{21}$ do not appear together in the same term of the symbol: the symbol can be separated into terms depending on one root, terms depending on the other, and terms depending on neither.

\vspace{-0pt}\section{Conclusions and Outlook}\vspace{-0pt}

In this work, we have computed a component of the two-loop eight-point NMHV amplitude in planar sYM at a specific kinematic point. We find that, while the individual integrals contributing to this amplitude do have letters depending on square roots of four-dimensional Gramians, these square roots cancel in the combination present in this component. In order to do this, we have employed an unusual direct integration strategy of breaking the integral into multiple integration pathways, and simplified our result from millions to thousands of terms using algebraic number theory.

This work shows that this particular component is free of square-root letters, but it does not establish that other components of the NMHV amplitude will not depend on such roots. In order to establish this, we would need to compute many more integrals, potentially of similar complexity. Alternatively, other methods may be able to compute these amplitudes much more efficiently, yielding a conclusive answer.

The use of symbol methods with square-root letters is still largely unexplored territory. While previous forays have involved heuristic or numerical elements (e.g.\ \mbox{\cite{Heller:2019gkq,Bonciani:2019jyb}}), our use of factorization in prime ideals should yield a more canonical and complete analysis of the relations between algebraic letters, and we believe similar methods should be applicable elsewhere.

It is interesting to ask if the cancellation of square roots we observed could have been detected at a later stage. For the individual integrals, better integration methods may exist that would make these cancellations manifest earlier, or even avoid the introduction of spurious roots altogether. For the full component amplitude, one might hope that some analog of Landau analysis might be possible.

\vspace{\fill}\vspace{-4pt}\section*{Acknowledgments}\vspace{-4pt}
We are grateful to Herbert Gangl and Francesco Moriello for discussion of alternate methods, to Mark Spradlin for motivation, and to Erik Panzer for help developing our integration strategy. We are also grateful to Song He for discussing soon to be published results \cite{SongHeToAppear} with close relevance to this work. This work was supported in part by the Danish Independent Research Fund under grant number DFF-4002-00037 (MV, MW), the Danish National Research Foundation (DNRF91), two grants from the Villum Fonden, a Starting Grant \mbox{(No.\ 757978)} from the European Research Council (JLB,AJM,MV,MvH,MW), the European Union's Horizon 2020 research and innovation program under grant agreement \mbox{No.\ 793151} (MvH), and a Carlsberg Postdoctoral Fellowship (CF18-0641) (AJM).
AJM, MV and MW thank the Center for the Fundamental Laws of Nature and JLB, AJM, MV and MW thank the Center Of Mathematical Sciences And Applications for kind hospitality during final stages of this work.
Finally, JLB, AJM and MvH are grateful for the hospitality of the Aspen Center for Physics, which is supported by National Science Foundation grant PHY-1607611.

\newpage\appendix

\vspace{-0pt}\section{A Proposal for Representing NMHV Octagon Amplitudes}\label{appendix:a_basis_for_nmhv_amplitudes}\vspace{-0pt}

In this appendix we describe a particular representation of eight-point NMHV amplitudes, analogous to the decomposition of hexagon and heptagon functions into specific bases. This is a bit outside the line of the main result in this work, but it does provide an independent logic behind why the particular component amplitude (\ref{component_integrand}) plays a special role. In order to do this, we must first introduce and motivate a small amount of new notation that we promise will be worthwhile.

\subsection{Notational Preliminaries: NMHV Yangian Invariants}

The reader should be aware that NMHV amplitudes can be expressed in terms of so-called $R$-invariants that, when expressed in momentum-twistor space, are superfunctions defined by
\eq{R[a,b,c,d,e]\equivR\frac{\delta^{1\times4}\big(\ab{bcde}\eta_a\pl\ab{cdea}\eta_b\pl\ab{deab}\eta_c\pl\ab{eabc}\eta_d\pl\ab{abcd}\eta_e\big)}{\ab{abcd}\ab{bcde}\ab{cdea}\ab{deab}\ab{eabc}}\,\label{r_invariant_defined}}
for any five (super-)momentum twistors labelled by $\{a,b,c,d,e\}$. An equivalent definition of the $R$-invariant is that it is simply the five-particle NMHV tree-level amplitude involving the momentum twistors $\{a,b,c,d,e\}$. It will turn out to be useful to consider NMHV tree-level amplitudes involving other sets of external particles including sets of more than five. In particular, let us use the symbol 
\eq{\amp{n}\equivR\r{1\cdots n}\equivR\mathcal{A}^{(k=1),L=0}_{n}(z_{\color{hred}1},\ldots,z_{\color{hred}n})\,\label{nmhv_tree_amplitude_notation}}
to denote the $n$-point NMHV tree-level amplitude involving momentum twistors $\{z_{\color{hred}1},\ldots,z_{\color{hred}n}\}$. (Recall that `$\mathfrak{A}$' is the Fraktur-script form of the letter `$A$'.) 
Thus, we may define the $R$-invariant simply as
\eq{R[1,2,3,4,5]\equivR\r{1,2,3,4,5}=\amp{5}\,.}

Especially at low multiplicity, we find it useful to denote tree amplitudes by which among the ambient $n$ twistors {\it they do not depend}. Because such notation, however convenient, is liable to cause confusion when several multiplicities are discussed, we propose to keep this information manifest in the way we write them. We denote these complements by
\eq{\rb{a\cdots b}{n}\equivR\r{[n]\backslash{\color{hblue}\{a,}{\color{hblue}\ldots,}{\color{hblue}b\}}}\,.\label{complementary_notation_defined}}
Notice that this would allow us to write
\eq{\amp{n}=\r{1\cdots n}\equivL\rb{\,}{n}}
---a notation that we cannot imagine ever actually using. More realistically, however, we should notice that in this notation the symbol for a single $R$-invariant would be multiplicity dependent. For example,
\eq{R[1,2,3,4,5]=\r{12345}=\rb{6}{6}=\rb{67}{7}=\rb{678}{8}=\cdots=\rb{6\cdots n}{n}\,.}

One (BCFW) representation (among many) of the NMHV tree amplitude (\ref{nmhv_tree_amplitude_notation}) would be,
\eq{\amp{n}=\amp{n-1}+\sum_{a=3}^{n-2}\r{1\,\,a\,\mi1a\,\,n\,\mi1\,n}=\sum_{a=3}^{n-2}\sum_{b=a+2}^{n}\r{1\,\,a\,\mi1a\,\,b\,\mi1\,b}\,;\label{bcfw_formula}}
but as already mentioned, we will have little recourse to decompose tree amplitudes into smaller objects. This is in part because, while $\r{1\cdots n}$ is in fact dihedrally-invariant in its indices, no {\it formula} of the form (\ref{bcfw_formula}) will make this manifest. 

Equivalence between various dihedrally-related BCFW formulae (\ref{bcfw_formula}) generates all the functional relations among $R$-invariants. In general, there are $\binom{n-1}{4}$ linearly independent $n$-point NMHV Yangian invariants.

At seven particles, for example, there are 15 linearly independent superfunctions into which any amplitude may be decomposed. Although $7$ does not divide 15, most authors (see e.g.\ \cite{Drummond:2008bq,Dixon:2016nkn,Drummond:2018caf}) have chosen to write heptagon functions in terms of the cyclic seeds $\{\rb{12}{7},\rb{14}{7},\amp{7}\}$ which generate 2 cyclic classes of length 7 and one cyclic singlet, $\amp{7}$. That is, these authors have chosen to decompose all \emph{other} 7-point superfunctions according to the `elimination rules' generated cyclically by
\eq{\begin{split}
\rb{13}{7}&=-\rb{34}{7}-\rb{56}{7}-\rb{71}{7}-\rb{36}{7}-\rb{51}{7}+\amp{7}\,,\\
\rb{1}{7}&=-\rb{34}{7}-\rb{56}{7}-\rb{36}{7}+\amp{7}\,.
\end{split}\label{heptagon_stupid_basis_eliminations}}
Having used such eliminations, the heptagon ratio function can be written as
\eq{R^L_{{\color{hred}7}}\equivL\Big[\Big(\rb{12}{7}V_{\rb{12}{7}}^{7,(L)}\pl\rb{14}{7}V_{\rb{14}{7}}^{7,(L)}\pl\amp{7}V_0^{7,(L)}\Big)\pl\text{cyclic}_7\Big]\,.}
(We believe that a \emph{better basis} for heptagon amplitudes would have been generated by $\{\rb{1}{7},\rb{12}{7},\amp{7}\}$, but this is not presently our concern.) Let us now describe a similar basis for eight-point NMHV Yangian invariants that is in a precise sense `optimal'.

\subsection{An Optimal Basis for Octagonal NMHV Amplitudes}

Unlike for seven particles (which is somewhat anomalously nice), there is no easy way to choose among the 56 different $R$-invariants---7 cyclic classes---into non-redundant classes spanning $35=\binom{7}{4}$ independent superfunctions. The situation is not obviously much improved if we include the cyclic singlet $\amp{8}$, or other lower-point tree-level amplitudes. Including also superfunctions corresponding to tree-level amplitudes involving intermediate subsets of the 8 legs, we have 13 cyclic classes of superfunctions, generated by 
\eq{\hspace{-80pt}\big\{\rb{123}{8},\!\rb{124}{8},\!\rb{125}{8},\!\rb{126}{8},\!\rb{134}{8},\!\rb{135}{8},\!\rb{136}{8},\!\rb{12}{8},\!\rb{13}{8},\!\rb{14}{8},\!\rb{15}{8},\!\rb{1}{8},\!\amp{8}\big\}\,.\hspace{-50pt}\label{eight_point_cyclic_reps}}
From this list, how are we to choose a basis of length 35? Of the cyclic classes generated by those in (\ref{eight_point_cyclic_reps}), all but two represent classes of length 8. The exceptions are $\amp{8}$ and $\rb{15}{8}\!=\!\r{234678}$, which forms a class of length 4. We are virtually forced to consider the inclusion of this length-4 class into our basis, as any other choice would lead to even greater redundancy.

Including $\amp{8}$, the four cyclic images of $\rb{15}{8}\!=\!\r{234678}$, and some other choice of four length-8 cyclic classes from among those generated by (\ref{eight_point_cyclic_reps}), we would have 37 superfunctions in all. In the best case, the two redundancies could be captured entirely by the length-four class (as 2 divides 4 nicely), with the rest independent. It turns out that there are 172 such choices available. The basis choice we describe presently is the one in which the `elimination rules' of all other superfunctions (in the sense of (\ref{heptagon_stupid_basis_eliminations})) involve the shortest expressions. 

The basis we propose can be defined first in terms of the 37 functions generated by the seeds
\vspace{2pt}\eq{\hspace{-130pt}\begin{array}{@{}l@{}l@{}l@{}l@{}l@{}l@{}}\fwbox{8.5pt}{\f{a}_1}\equivR\r{12345}=\rb{678}{8}\,,\;\;\fwbox{8.5pt}{\f{b}_1}\equivR\r{12346}=\rb{578}{8}\,,\;\;&\fwbox{8.5pt}{\f{c}_1}\equivR\r{123456}=\rb{78}{8}\,,\\
\fwbox{8.5pt}{\f{d}^0_1}\equivR\r{123567}=\rb{48}{8}\,,\;\;\fwbox{8.5pt}{\f{e}_1}\equivR\r{1234567}=\rb{8}{8}\,,\;\;&\fwbox{8.5pt}{\f{f}}\equivR\r{12345678}\hspace{-0.75pt}=\hspace{-0.75pt}\amp{8}\,,\end{array}\hspace{-100pt}\label{eight_point_basis}}
with other basis elements generated by cyclic rotations. Before we discuss the final, non-redundant basis, it is worthwhile to enumerate the (cyclic generators of all) elimination rules---by which non-basis superfunctions may be expanded:
\vspace{-4pt}\eq{\hspace{-70pt}\begin{array}{@{}l@{}l@{}l@{}}\rb{124}{8}&=\mi\rb{467}{8}\pl\rb{12}{8}\pl\rb{67}{8}\pl\rb{4}{8}\mi\amp{8}&=\mi\f{b}_8\pl\f{c}_3\pl\f{c}_8\pl\f{e}_5\mi\f{f}\,;\\
\rb{125}{8}&=\mi\rb{123}{8}\mi\rb{127}{8}\pl\rb{12}{8}&=\mi\f{a}_4\mi\f{b}_3\pl\f{c}_3\,;\\
\rb{126}{8}&=\mi\rb{128}{8}\pl\rb{467}{8}\mi\rb{67}{8}\mi\rb{4}{8}\pl\amp{8}&=\mi\f{a}_3\pl\f{b}_8\mi\f{c}_8\mi\f{e}_5\pl\f{f}\,;\\
\fwboxR{0pt}{*}\rb{135}{8}&=\phantom{\mi}\rb{178}{8}\pl\rb{567}{8}\pl\rb{15}{8}\mi\rb{3}{8}\mi\amp{8}&=\phantom{\mi}\f{a}_2\pl\f{a}_8\pl\f{d}^0_2\pl\f{e}_4\mi\f{f}\,;\\
\rb{136}{8}&=\phantom{\mi}\rb{567}{8}\mi\rb{134}{8}\pl\rb{356}{8}\mi\rb{18}{8}\mi\rb{56}{8}\pl\rb{1}{8}&=\phantom{\mi}\f{a}_8\mi\f{b}_5\pl\f{b}_7\mi\f{c}_2\mi\f{c}_7\pl\f{e}_2\,;\\
\rb{13}{8}&=\phantom{\mi}\rb{567}{8}\pl\rb{1}{8}\pl\rb{3}{8}\mi\amp{8}&=\phantom{\mi}\f{a}_8\pl\f{e}_2\pl\f{e}_4\mi\f{f}\,;\\
\rb{14}{8}&=\phantom{\mi}\rb{134}{8}\mi\rb{467}{8}\mi\rb{34}{8}\pl\rb{4}{8}&=\phantom{\mi}\f{b}_5\mi\f{b}_8\mi\f{c}_5\pl\f{e}_5\,;\\
\fwboxR{0pt}{\f{d}_3^0\!\!=}\rb{26}{8}&=\mi\rb{678}{8}\mi\rb{128}{8}\mi\rb{234}{8}\mi\rb{456}{8}\mi\rb{48}{8}\pl\amp{8}\,&=\mi\f{a}_1\mi\f{a}_3\mi\f{a}_5\mi\f{a}_7\mi\f{d}_1^0\pl\f{f}\,;\\
\fwboxR{0pt}{\f{d}_4^0\!\!=}\rb{37}{8}&=\mi\rb{178}{8}\mi\rb{123}{8}\mi\rb{345}{8}\mi\rb{567}{8}\mi\rb{15}{8}\pl\amp{8}\,&=\mi\f{a}_2\mi\f{a}_4\mi\f{a}_6\mi\f{a}_8\mi\f{d}_2^0\pl\f{f}\,.
\end{array}\hspace{-50pt}\label{eight_point_elimination_rules}}
There are a few things to note about these decompositions. As always, other superfunctions are eliminated according to rotations of (\ref{eight_point_elimination_rules}). In addition, there are two aspects of (\ref{eight_point_elimination_rules}) regarding $\f{d}_i^0$ that deserve comment. First, note that the only superfunction from (\ref{eight_point_basis}) whose decomposition involves $\f{d}_i^0$ (except those of the $\f{d}_i^0$'s) is $\rb{135}{8}$---indicated with a `$*$' in (\ref{eight_point_elimination_rules}).\footnote{It is worth mentioning that this particular superfunction, $\rb{135}{8}$, does not appear as any leading singularity (hence integral coefficient) until at three loops---where it certainly appears.}

The second aspect to notice about the elimination rules (\ref{eight_point_elimination_rules}) is that the last two are for $\f{d}_3^0$ and $\f{d}_4^0$, which are generated by our initial seeds upon rotation. As evidenced by the simple fact that they have elimination rules (and also that $35\!=\!37\mi2$), these two will not be basis elements. Moreover, it is easy to see that
\eq{\f{d}_1^0\pl\f{d}_3^0=\f{f}\mi\f{a}_1\mi\f{a}_3\mi\f{a}_5\mi\f{a}_7\quad\text{and similarly,}\quad \f{d}_2^0\pl\f{d}_4^0=\f{f}\mi\f{a}_2\mi\f{a}_4\mi\f{a}_6\mi\f{a}_8\,.}
However, the differences between them are good basis elements. And up to the alternating sign, they form a length-2 cyclic class of superfunctions. Let us define
\eq{\f{d}_1\equivR\f{d}_1^0\mi\f{d}_3^0\quad\text{and}\quad\f{d}_2\equivR\f{d}_2^0\mi\f{d}_4^0\,.}
These, combined with the other basis elements in (\ref{eight_point_basis}), non-redundantly span the space of 35 independent superfunctions in terms of four cyclic classes of length 8, one of length 2, and one of length 1. This is our proposed basis for eight-point NMHV amplitudes. 

In this basis, the eight-point NMHV ratio function may be represented as
\eq{R^{(L)}_{{\color{hred}8}}\equivR\Big[\Big(\f{a}_1V_{\f{a}}^{(L)}\pl\f{b}_1V_{\f{b}}^{(L)}\pl\f{c}_1V_{\f{c}}^{(L)}\pl\f{d}_1V_{\f{d}}^{(L)}\pl\f{e}_1V_{\f{e}}^{(L)}\pl\f{f}V_{\f{f}}^{(L)}\Big)\pl\text{cyclic}_8\Big]\,.}
(As with seven points, please notice that we are adding all of these terms \mbox{(8-fold-)} cyclically. This has the admittedly unfortunate effect of causing $V_{\f{f}}^{(0)}$ to be $1/8$; it will also require us to account for the over-counting in $\smash{V_{\f{d}}^{(L)}}$.)

For reference, at one loop, these are easy to write explicitly \cite{Elvang:2009ya,Bourjaily:2013mma}. They are
\eq{\begin{split}
V^{(1)}_{\f{a}}=\,&\mi\Li{2}{1\mi v_2}\mi\Li{2}{1\mi u_1u_4v_4}\mi\log(u_2)\log(u_3)\mi\log(u_1u_4v_4)\log(v_2)\pl\Zeta{2}\,,\\
V^{(1)}_{\f{b}}=\,&\phantom{+}\Li{2}{1\mi u_5v_1}\mi\Li{2}{1\mi u_2u_5v_1}\mi\Li{2}{1\mi u_4v_3}\pl\Li{2}{1\mi u_4u_7v_3}\\
&\mi\log(u_2)\log(u_4v_3)\pl\log(u_5v_1)\log(u_7)\,,\\
\hspace{-140pt}V^{(1)}_{\f{c}}=\,&\mi\Li{2}{1\mi u_7}\mi\Li{2}{1\mi u_5v_1}\pl\Li{2}{1\mi u_2u_5v_1}\mi\Li{2}{1\mi u_2v_2}\pl\Li{2}{1\mi u_2u_7v_2}\hspace{-100pt}\\
&\mi\log(u_4v_3)\log(v_2)\mi\log(u_5v_1)\log(u_7)\,,\\
V^{(1)}_{\f{d}}=\,&\phantom{+}0\,,\\
V^{(1)}_{\f{e}}=\,&\mi\Li{2}{1\mi u_2u_7v_2}\mi\Li{2}{1\mi u_8v_4}\mi\log(u_2u_7v_2)\log(u_8v_4)\pl\Zeta{2}\,,\\
\hspace{-120pt}V^{(1)}_{\f{f}}=\,&\phantom{+}\Li{2}{1\mi u_1}\pl\frac{1}{2}\Li{2}{1\mi v_1}\pl\Li{2}{1\mi u_1v_1}\mi\frac{1}{2}\log(v_1)\log(v_2)\pl\frac{3}{4}\log(v_1)\log(v_3)\hspace{-100pt}\\
&\;\pl\log(u_1v_4)\log(u_2u_3v_3)\mi\Zeta{2}\,.
\end{split}\label{one_loop_octagon_function}}
We have written these function in terms of the 12 multiplicatively independent dual-conformally invariant cross-ratios,
\eq{u_1\equivR\u{13}{48},\quad v_1\equivR\u{14}{58}\quad\text{with}\quad u_i\equivR r^{(i-1)}(u_1),\quad v_i\equivR r^{(i-1)}(v_1)\,.}
Notice that $V_\f{d}$ is zero at one loop. At two loops, it is not hard to confirm that
\eq{V_{\f{d}}^{(2)}=-\frac{1}{4}\left[\!\!\octagonk-\octagonkPrime\!\!\right]\,.}

\newpage
\vspace{-0pt}\section{Some Notions of Algebraic Number Theory}\vspace{-0pt}\label{sec:alg_num_th}
%
When working with symbols, it is valuable to be able to put them into a
canonical form, for instance to decide whether two symbols are equal. 
As an example, many of the amplitudes that have been computed in planar sYM to date can be uniquely expressed in terms of a known set of Pl\"ucker coordinates. 
In more complicated amplitudes, a basis of symbol letters is not generally known. In such cases, we can simply factorize each
symbol letter, as long as this factorization is unique. 

It is easy to see that factorization will give rise to a unique expression when all symbol letters are integers. However, this is not automatic once algebraic roots are introduced. Consider, for instance, the situation where $\sqrt{-5}$ appears in some letters. The number 9 then has two `factorizations': 
\eq{9 = 3 \times 3 = (2 + \sqrt{-5})(2 - \sqrt{-5})\,,}
where the second factorization of $9$ is possible when viewed as an element of \(\mathbb{Z}[\sqrt{-5}]\). By \(\mathbb{Z}[\sqrt{-5}]\), we denote the set of numbers of
type \(a + b \sqrt{-5}\) for \(a, b \in \mathbb{Z}\), with the obvious
addition and multiplication properties.\footnote{We should not think of \(\sqrt{-5}\) as being a complex
  number, but rather as an abstract symbol whose property is that it
  squares to \(-5\).  In fact, \(\mathbb{Z}[\sqrt{-5}]\) can be
  embedded in the complex numbers in two ways, by sending
  \(\sqrt{-5}\) to each of the two roots of \(-5\) in \(\mathbb{C}\).} This set of numbers, with
these operations, defines a \emph{ring}.

From the example above it looks like \(9\) can be factorized in two
different ways, but perhaps unique factorization can still be salvaged
if some of the factors can be further factored. It turns out that this is not what is happening here.

Before clarifying what is happening, we need to make a distinction
between irreducible and prime elements of a ring \(R\).  First, we
introduce the notion of \emph{unit}.  The elements of \(R\) which have
multiplicative inverses are the units of \(R\) (denoted by \(U(R)\)).
For the integers, the units are \(\pm 1\).  An element \(x \in R\) is
\emph{irreducible} if it can not be written as a product of two elements of
\(R\) neither of which is a unit.  Finally, an element \(x \in R\) is
\emph{prime} if for any \(a, b \in R\) such that \(x\) divides
\(a b\), then it divides \(a\) or \(b\).  For the integers there is no
distinction between primes and irreducibles, but in general rings
there is.

We now return to the above example: is \(3\) a prime in
\(\mathbb{Z}[\sqrt{-5}]\)?  We can show that it is not.  If it were
prime, it would follow from the fact that \(3\) divides \((2 + \sqrt{-5}) (2 - \sqrt{-5})\) that it also divides either \(2 + \sqrt{-5}\) or
\(2 - \sqrt{-5}\).  But \(3\) divides \(a + b \sqrt{-5}\) only if it
divides both \(a\) and \(b\), which is not the case here.

Is \(3\) irreducible instead?  One can show that the units of
\(\mathbb{Z}[\sqrt{-5}]\) are \(\pm 1\).  It is then a simple exercise
to show that \(3\) is indeed irreducible (just use the definition and
show that there are no suitable solutions).  So the hope that perhaps
each of the terms in the factorization can be factorized further to a
prime decomposition which is the same in the LHS and RHS is not
fulfilled.  We conclude that \(\mathbb{Z}[\sqrt{-5}]\) is not a UFD
(unique factorization domain).

For this reason, it may look like there is no way to achieve unique factorization. But if we enlarge our perspective a little, we can recover
this desired property.  We will now explain how to do this.  The
construction we will describe is possible for rings which are
\emph{Dedekind domains}.

Let us start with the familiar case of integers.  In this case, to a
prime \(p\) we associate the set of all its multiples.  This set has
two important properties.  First, it is closed under addition;
second, multiplying it by any integer lands us back in the same set.
This is just the definition of an \emph{ideal} of the ring of integers
\(\mathbb{Z}\).  For the case of a prime we obtain a \emph{prime
  ideal}, but the construction works in general.  The set of multiples
of \(p\) is denoted by \((p)\).  This is also called the ideal
generated by \(p\).

The notion of divisibility can be translated to the language of
ideals: we say that \(a\) divides \(b\) if \((b) \subseteq (a)\).  It
is easy to check that this corresponds to the usual notion of
divisibility for the integers.  Now that we have expressed
divisibility in terms of ideals, we may consider ideals generated by
more than one element.  The ideals generated by one element, such as
\((p)\), are called \emph{principal ideals}.  An ideal generated by
two elements \(a\) and \(b\) is denoted by \((a, b)\); as a set,
it contains the linear combinations \(m a + n b\) where \(m\), \(n\)
belong to the ring and \(a\), \(b\) belong to the ideal.  This
satisfies all the properties of an ideal.

Ideals can be multiplied; we have \((p) (q) = (p q)\) and \((a, b)(c, d)
= (a c, a d, b c, b d)\) and the pattern continues in the obvious way,
for ideals generated by more generators.  These ideals have some pretty
obvious properties:
\eq{(a, b) = (a \pm b, b)\,,\qquad(a, b, a \pm b) = (a, b)\,,\qquad (1, a) = (1)\,.}
Using these rules we can compute the following products, which will be
useful momentarily:
\begin{align}
\hspace{-15pt} (3, 1 + \sqrt{-5}) (3, 1 - \sqrt{-5}) &=(9, 3 + 3 \sqrt{-5}, 3 - 3 \sqrt{-5}, 6) =(9, 3 + 3 \sqrt{-5}, 6) \\
&=(3) (3, 1 + \sqrt{-5}, 2) = (3) (1, 1 + \sqrt{-5}, 2) =(3) (1) = (3).\nonumber
\end{align}
Similarly, we find
\begin{align}
\hspace{-8pt}(3, 1 \pl \sqrt{-5})^2 &=(9, 3 \pl 3 \sqrt{-5}, \mi4 \pl2 \sqrt{-5}) =(9, \mi6 \pl 3 \sqrt{-5}, \mi4\pl 2 \sqrt{-5})\nonumber\\
& = ((2 \pl \sqrt{-5}) (2 \mi \sqrt{-5}), \mi3 (2 \mi\sqrt{-5}), \mi4 (2 \mi \sqrt{-5}))\\
&=(2 \mi \sqrt{-5}) (2 \pl \sqrt{-5}, \mi3, \mi4) =(2 \mi \sqrt{-5}) (2 \pl \sqrt{-5}, 1, \mi4) =(2 \mi \sqrt{-5}).\nonumber
\end{align}
We also have \((3, 1 + \sqrt{-5})^2 = (2 + \sqrt{-5})\).

Now that we have made the transition from elements of a ring to the
principal ideal they generate, we can explain the change of
perspective mentioned above.  Instead of considering principal ideals,
we consider ideals generated by any number of generators.  Indeed, now
we can refine the factorization as follows:
\begin{equation}
    (9) = (3) (3) = (2 + \sqrt{-5}) (2 - \sqrt{-5}) =
    (3, 1 + \sqrt{-5})^2 (3, 1 - \sqrt{-5})^2.
\end{equation}
To finish, we should show that the ideals appearing in this
factorization are prime.  We will not do this explicitly here.

This works in general.  The factorization is unique in the following
sense: any ideal can be decomposed as a product of prime ideals, up to
ordering.  Finally, we have achieved unique factorization, but at the
cost that the factors are some abstract, less familiar quantities.

An \emph{algebraic number field} is a finite extension of
\(\mathbb{Q}\) constructed as follows.  Consider a root \(\polyalpha\) of
a degree \(n\) polynomial with rational coefficients.  Then,
\(\mathbb{Q}[\polyalpha]\) is the ring generated by rational linear
combinations of powers \(0\) through \(n - 1\) of \(\polyalpha\) (higher
powers can be reduced).  We also define \(K = \mathbb{Q}(\polyalpha)\) as
the field generated by \(\polyalpha\) (whose elements are ratios of
elements of \(\mathbb{Q}[\polyalpha]\)).  Inside \(K\) we find the
\emph{algebraic integers} \(\mathcal{O}_K\) which are the elements of
\(K\) whose minimal polynomial is monic\footnote{A \emph{monic}
  polynomial has its leading coefficient equal to one.} and with
integer coefficients.  It is a theorem that the ring of algebraic
integers \(\mathcal{O}_K\) of an algebraic number field \(K\) is a
Dedekind domain, so it has a unique factorization.

Some of the letters we would like to factorize are not actually
algebraic integers, so we cannot construct an ideal they generate
inside \(\mathcal{O}_K\).  Nevertheless, we can construct a
\emph{fractional ideal} instead, which is a slight generalization of
the notion of ideal.  We will not give a full definition here, but the
reader who wants to have an intuition for what a fractional ideal is
can think of \(\frac p q \cdot \mathbb{Z}\) as a fractional ideal of
\(\mathbb{Z}\).  In other words, we also allow denominators.

Now the strategy for computing relations between several elements of a
number field \(K\) should be clear.  For each of these elements we
compute the prime ideal decomposition of the principal fractional ideal
they generate.  The exponents form a matrix with integer coefficients whose
rows are labeled by the elements of \(K\) and whose columns are
labeled by the prime ideals.  Every element of the left kernel of this
matrix yields a multiplicative relation between the given elements of
\(K\).

Historically, it was Kummer who started developing these ideas in
connection with Fermat's conjecture.  His ideas were refined and
generalized by Dedekind, Hilbert, Noether and many others. A good reference and resource for the material described in this appendix is~\cite{Stein:2012}.

\newpage
\vspace{-10pt}
\providecommand{\href}[2]{#2}\begingroup\raggedright\endgroup
\end{document}